\documentclass{article}
\usepackage{graphicx} 
\usepackage{mathptmx}
\usepackage{latexsym}
\usepackage[numbers,sort&compress]{natbib}
\usepackage[colorlinks,citecolor=blue,urlcolor=blue,linkcolor=blue]{hyperref}
\usepackage{amsmath}
\usepackage{exscale} 
\usepackage{authblk}
\title{Thermodynamics of FLRW universe in Quadratic Gravity}
\author[1]{Navid Safarzadeh Ilkhchi\thanks{Email: \texttt{navidsafarzadeh1402@ms.tabrizu.ac.ir}}}
\author[2]{Amin Rezaei Akbarieh\thanks{Email: \texttt{am.rezaei@tabrizu.ac.ir}}}
\author[3]{Yaghoub Heydarzade\thanks{Email: \texttt{yheydarzade@bilkent.edu.tr}}}

\affil[1,2]{Faculty of Physics, University of Tabriz, Tabriz, 51666-16471, Iran }
\affil[3]{Department of Mathematics, Faculty of Sciences, Bilkent University, 06800 Ankara, Turkey}
\date{July 2025}

\begin{document}
\maketitle

\begin{abstract}
In this paper, we investigate the thermodynamic aspects of quadratic gravity in a $D$-dimensional Friedmann-Lemaitre-Robertson-Walker (FLRW) universe. First, we derive the field equations and the effective energy-momentum tensor for quadratic gravity. Then, using these equations, we obtain the generalized Misner-Sharp energy within the framework of this model. We consider the thermodynamic behavior of the apparent horizon and derive the equations of state related to the pressure, temperature, and radius of the apparent horizon. Using the thermodynamic pressure, we obtain the critical points corresponding to phase transitions. We determine the critical temperature and critical radius in terms of model parameters, including the quadratic coupling and the cosmological constant. We also examine key thermodynamic quantities, such as Wald entropy, specific heat at constant pressure, enthalpy, Gibbs free energy. By examining the behavior of these quantities, we can gain insight into the thermodynamic stability of the quadratic gravity model. In particular, we find that quadratic terms change the stability conditions and can lead to new thermodynamic behaviors compared to general relativity. 
\end{abstract}

\section{Introduction}
In recent decades, serious efforts have been made to obtain a consistent and comprehensive theory of gravity at low and high energy scales, and it can almost be said that it has been one of the main topics of modern theoretical physics. Although general relativity (GR) is a very successful theory in describing gravitational phenomena \cite{Clifton:2011jh,Berti:2015itd}, it faces several challenges in different regimes, including the region near singularities, the early universe, and at the interface with quantum theory \cite{Myung:2022biw,Kaneta:2025dcs}. To overcome these challenges, many solutions have been proposed, including the development of modified theories of gravity in which the standard Einstein-Hilbert interaction is generalized with expressions related to high curvatures or quantum corrections \cite{Sotiriou:2008rp,DeFelice:2010aj,Amelino-Camelia:2003ezw,Sanyal:2001ws}.\\
One of the interesting models in the field of modified gravity is quadratic gravity, which has been the subject of long-term attention \cite{Salvio:2019ewf,Donoghue:2021cza,Gurses:2024tka}. In addition to the standard Einstein-Hilbert term, the quadratic gravity model includes correction terms such as the square of the Ricci scalar ($R^2$), the square of the Ricci tensor ($R_{\mu\nu}R^{\mu\nu}$), and the square of the Riemann tensor ($R_{\mu\nu\rho\sigma}R^{\mu\nu\rho\sigma}$). It should be noted that these correction terms, which include higher-order curvatures, naturally appear in the effective interactions in quantum gravity theories such as string theory, and loop quantum gravity \cite{Mohaupt:2005jd,Alonso-Serrano:2024amg,Cecotti:2014ipa,Ashtekar:2011ni}. By considering these quadratic modifications, we can study the quantum modifications of the classical gravitational interaction, and as a result, this allows us to describe gravity at high energy more completely and accurately \cite{Bojowald:2006ww,Date:2008gq}.\\
Quadratic gravity has been studied in various fields, from cosmology and the thermodynamics of black holes to the study of higher-dimensional holographic spaces \cite{CANTATA:2021asi,Lu:2015cqa}. In \cite{Donoghue:2021cza} it was shown that this model can be a good candidate for quantum gravity. By investigating the linear stability of second-order gravity in the high-frequency geometric optics approximation, it was found that both types of perturbations—scalar modes and gravitational modes—propagating on spherically symmetric and axially symmetric solutions are linearly stable. In both the scalar-dominant and gravitational-wave-dominant regimes, the dispersion relations do not lead to exponential growth, indicating the linear stability of the theory in these parts \cite{Ayzenberg:2013wua}. It has been shown that adding quadratic curvature corrections to gravitational theories can lead to dynamical stability of isotropic cosmological solutions in the early stages of the universe, indicating a key role of these corrections in pushing the dynamics of the universe towards isotropy near the early singularity \cite{Barrow:2006xb,Barrow:2007pm}. Also, these theories can well explain different epochs of the universe, such as inflation, the period of radiation dominance, and the late time acceleration that indicates the key role of this kind of gravity in understanding the structure and evolution of the universe \cite{Chakraborty:2021enp,BeltranJimenez:2016wxw}. Additionally, it was shown that the pure $R^2$ theory is ghost-free and on flat backgrounds, it propagates only a massless scalar mode \cite{Alvarez-Gaume:2015rwa}.\\
Since the thermodynamics of gravitational systems has an important role in clarifying the fundamental nature of spacetime and gravity, in the present work, we aim to examine the thermodynamics of quadratic gravity in FLRW universe. In GR, fundamental work on the thermodynamics of black holes—in particular, the discovery of the four laws of black hole mechanics and Hawking radiation—has revealed deep connections between the geometry of spacetime, quantum field theory, and thermodynamic behavior  \cite{Sebastiani:2023brr,Padmanabhan:2015zmr,Bamba:2011jq}. These results can strengthen the conjecture that gravity itself may be an emergent phenomenon and that thermodynamic principles serve as a way to understand the spacetime underpinning it. The extension of these thermodynamic ideas to modified theories of gravity, such as quadratic gravity, is a powerful tool for considering the implications of higher-order modifications in understanding the dynamics of these models. Several studies have investigated the thermodynamic properties of quadratic gravity in various situations, such as black hole solutions in higher-derivative gravity theories and the effect of quadratic modifications on their entropy and stability \cite{Sen:2005wa,Lu:2015cqa,Camps:2013zua,Gurses:2024tka}. In the context of the FLRW universe, thermodynamic studies have shown that quadratic curvature terms can significantly alter the properties of the apparent horizon, such as its temperature and thermodynamic pressure, influencing the cosmological evolution in high-curvature regimes \cite{Saavedra:2023rfq}. These findings emphasize the importance of gravitational modifications in understanding the thermodynamic behavior of the universe, particularly in higher dimensional spacetimes \cite{Saavedra:2023rfq}. It has been shown that quadratic terms can lead to significant deviations from the familiar behavior of GR, including modified expressions for the entropy of the horizon, modified stability conditions, and new phases of gravitational systems \cite{Bellucci:2009nn}.\\
Despite significant research in the field of quadratic gravity, there is still a need for systematic and comprehensive investigations of its thermodynamic properties in the cosmic background, especially in higher-dimensional spacetimes \cite{CANTATA:2021asi}. Such studies are essential to understand how higher-order curvature modifications affect the thermodynamic behavior of the universe and may contribute to a better understanding of the microscopic degrees of freedom that underline gravitational dynamics. In this paper, we aim to study the thermodynamics of quadratic gravity in detail in the background of the $D$-dimensional FLRW universe. We first study the generalized action of quadratic gravity, which includes the second-order curvature terms with the coupling parameters $\alpha$, $\beta$, and $\gamma$ as well as the cosmological constant $\Lambda$. We then derive the field equations and the effective energy-momentum tensor that include the contributions of higher-order terms. One of the main aspects of our study is the generalization of the Misner-Sharp energy in $D$ dimensional spacetime within the framework of the proposed model. This quasi-local energy measure plays a fundamental role in connecting the gravitational field equations with the thermodynamic laws and is an essential tool in understanding the connection between geometry and thermodynamics in various theories of gravity. We then derive the conditions under which the Misner-Sharp energy is well-defined, and we also show that the energy obtained reduces to the familiar results of GR as long as we do not consider the second-order terms.\\
Using this approach, we study how phase transitions and critical points appear in the thermodynamic description of the apparent horizon in the $D$-dimensional FLRW universe. We show how the thermodynamic equation relates the pressure, temperature, and radius of the apparent horizon. We obtain the expressions for the critical temperature and critical radius in terms of the parameters of the theory, such as the quadratic coupling constants and the cosmological constant. Furthermore,  other important thermodynamic quantities, such as the Wald entropy, specific heat, and Gibbs free energy, are obtained to show how the quadratic gravity modifies these properties. Finally, we to show discuss how these results relate to the thermodynamic behavior and stability of the system compared to general relativity.\\
This paper is organized as follows: In Section \ref{sec2}, we introduce the theoretical framework of quadratic gravity and derive the associated field equations, including the explicit form of the effective energy-momentum tensor. In Section \ref{sec3}, we present the construction of the generalized Misner-Sharp energy and discuss the necessary conditions for its validity. In Section \ref{sec4}, we investigate the conditions for criticality and analyze the phase transitions that emerge in this framework. Section \ref{sec5} is devoted to computing the essential thermodynamic quantities and discussing their implications for the system’s thermodynamic stability and phase behavior. Finally, in Section \ref{sec6}, we summarize our main findings and discuss their broader implications for the study of modified gravity theories and the thermodynamic properties of the universe.

\section{The Quadratic Gravity}\label{sec2}
In this section, we derive the equations of motion for quadratic gravity and compute the energy-momentum tensor components in the $D$-dimensional FLRW background. The action of quadratic gravity is given by \cite{Gurses:2024tka}
\begin{eqnarray}\label{201}
S&=&\int d^Dx\sqrt{-g}\bigg[ \frac{1}{\kappa} (R-2\Lambda) + \alpha R^2 + \beta R^2_{\mu \nu}+ \gamma (R^2_{\mu \nu \sigma \rho} - 4R^2_{\mu \nu} + R^2) + \mathcal{L}_M \bigg] ,\;\;\;\;\;\;\
\end{eqnarray}
where $\kappa=8\pi G_D$ (in which $G_D$ represents
the Newton’s gravitational constant in $D$-dimensions), $\Lambda$ is the cosmological constant, $R$ stands for Ricci scalar, $R_{\mu\nu}$ is Ricci tensor, $R_{\mu\nu\sigma\rho}$ is Riemann tensor, and also the constants $\alpha$, $\beta$ and $\gamma$ are coupling parameters of the quadratic terms, while $\mathcal{L}_M$ represents the matter Lagrangian.\\
By varying the action (\ref{201}) with respect to the metric, we obtain the field equation
\begin{eqnarray}\label{202}
    \frac{1}{\kappa}\big(R_{\mu \nu} - \frac{1}{2}g_{\mu \nu} R + \Lambda g_{\mu \nu} \big)+ \mathcal{E}_{\mu \nu} = T_{\mu\nu},
\end{eqnarray}
where $T_{\mu\nu}$ is the matter energy-momentum tensor of a perfect fluid. $\mathcal{E}_{\mu\nu}$ includes the contribution of second-order curvature terms as
\begin{eqnarray}\label{203}
\mathcal{E}_{\mu\nu} &\equiv& 2\alpha R ( R_{\mu\nu} - \frac{1}{4} g_{\mu\nu} R )
+ (2\alpha + \beta) \left( g_{\mu\nu} \Box 
- \nabla_{\mu} \nabla_{\nu} \right) R \nonumber \\
&&+ 2\gamma \big[ R R_{\mu\nu} - 2 R_{\mu\rho\nu\sigma} R^{\rho\sigma}
+ R_{\mu\sigma\rho\tau} R_{\nu}^{\ \sigma\rho\tau}
- 2 R_{\mu\rho} R_{\nu}^{\ \rho} \nonumber\\
&&- \frac{1}{4} g_{\mu\nu} \left( R_{\tau \lambda \sigma \rho}^2
- 4 R_{\sigma \rho}^2 + R^2 \right) \big]+ \beta \Box ( R_{\mu\nu} - \frac{1}{2} g_{\mu\nu} R )
 \nonumber\\
&&+ 2\beta ( R_{\mu\sigma\nu\rho} - \frac{1}{4} g_{\mu\nu} R_{\sigma\rho}) R^{\sigma\rho},
\end{eqnarray}
where $\Box = \nabla^{\mu} \nabla_{\mu}  $ denotes the d’Alembertian operator. In \cite{Gurses:2020kpv} it was shown that this tensor has a perfect fluid form for the FLRW metric.\\
The line element of an arbitrary $D$-dimensional FLRW universe is 
\begin{eqnarray}\label{204}
    ds^2=-dt^2+a^2(t)\left[\frac{dr^2}{1-kr^2}+r^2d\Omega^2_{D-2}\right],
\end{eqnarray}
where $a(t)$ is the scale factor, $k$ determines the spatial curvature, and $d\Omega^2_{D-2}$ represents the line element on a $(D-2)$-dimensional sphere. By considering metric (\ref{204}) and using (\ref{202}), we can obtain the temporal and radial components of the energy-momentum tensor, respectively, as follows 
\begin{eqnarray}\label{205}
\hspace{-4cm}T_t^{\;t} &=& \frac{ \Lambda}{\kappa}  - \frac{1}{2\kappa D a^4}\big[D(D-2)(D-1) \left( k + \dot{a}^2 \right) a^2  \nonumber \\
&& +  \kappa\sigma(D-1)  \left( k + \dot{a}^2 \right)^2 - \kappa(1-D) \left( 4(D-1)\alpha + D\beta \right)  \nonumber\\
&&\times \big( k^2 (D-2)^2 - (D^2-4) \dot{a}^4 - D a^2 \ddot{a}^2  \nonumber \\
&&  + 2 \dot{a}^2 \left( -2k(D-2) + (D-3)D a \ddot{a} \right) + 2 D a^2 \dot{a}\; \dddot{a} \big)\big],\nonumber
\\
\hspace{-22 cm}T_r^{\;r} &=&  \frac{ \Lambda}{\kappa}+ \frac{1}{2 \kappa D a^4} \Big[ \kappa  \sigma\left( k + \dot{a}^2 \right) \left[ - (D-5) \left( k + \dot{a}^2 \right) - 4 \ddot{a} \right]  \nonumber\\
&& + D a^2 \left[ -(D-3)(D-2) \left( k + \dot{a}^2 \right) - 2 (D-2) a \ddot{a} \right] \nonumber \\
&& - \kappa\left[ 4(D-1)\alpha + D\beta \right]  \Big\{ k^2 (D-5)(D-2)^2 \nonumber\\
&&- (D-5)(D-2)(2+D) \dot{a}^4  + 2 \dot{a}^2 \left( -2k (D-5)(D-2) \right.\nonumber\\
&&\left. + \left( 8 + D(12 + (D-9)D) \right) a \ddot{a} \right) +\left. a \ddot{a} \left( -8k (D-2)\right. \right.\nonumber \\
&& \left.+ 3(D-3)D a \ddot{a} \right) + 4 (D-3)D a^2 \dot{a} \dddot{a} + 2 D a^3 \ddddot{a} \Big\} \Big],\nonumber\\
\end{eqnarray}
where dot sign represents the derivative with respect to time, and $\sigma$ is defined as 
\begin{eqnarray}\label{206}
    \sigma\equiv ( D-4 ) ( D-2 ) \left[ (D-2)(D-1)\alpha + (D-3) D \gamma\right].\;\;\;\
\end{eqnarray}
We can eliminate the contribution of the higher derivatives of the scale factor, such as $\dddot{a}$ and $\ddddot{a}$, to the components of the energy-momentum tensor by appropriate choices of the parameters $\alpha$, $\beta$ and $\gamma$. From equation (\ref{205}), we can see that although the higher derivatives of the scale factor do not appear when $4(D-1)\alpha + D\beta=0$, there are still contributions of the quadratic terms. This choice is important because, in various gravitational models, higher-order derivatives may be linked to unstable degrees of freedom or the presence of ghosts in the theory. By eliminating these higher-order derivatives, we can ensure that the quadratic model becomes an effective gravitational theory at low energy scales, effectively mimicking the behavior of GR. In addition, eliminating these higher derivatives causes the cosmological dynamics, i.e., the FLRW equations of motion, to be limited to only the second-order derivatives. This makes the behavior of the universe in this model very similar to GR, without introducing unstable and unusual dynamics in the early universe \cite{Pravda:2016fue,Gurses:2024tka}.

\section{The Misner-Sharp Energy}\label{sec3}
In this section, we aim to find the Meisner-Sharp energy in a $D$-dimensional FLRW spacetime in the context of quadratic gravity. This energy is a quasi-local quantity that relates the gravitational field equations to the thermodynamic laws, and it can be defined only if the convergence condition is satisfied. It can be shown that in FLRW space, the Misner–Sharp energy is equivalent to the sum of the effective energy inside a sphere of a given radius; this energy includes both the matter contribution and the effects of spacetime curvature. Therefore, using Misner-Sharp energy facilitates the thermodynamic analysis of the apparent horizon and gravitational processes \cite{Cai:2009qf,Akbarieh:2025vau}.\\
We start by considering the unified first law in the form of the differential of the
effective Misner-Sharp energy, which can be expressed as \cite{Hayward:1998ee,Cai:2009qf}
\begin{eqnarray}\label{307}
    dE_{MS} =  \mathbf{A} \Psi_a dx^a + WdV,
\end{eqnarray}
where $ \mathbf{A} = V^1_{D-2} R^{D-2}$ is the area of the $(D-2)$-dimensional sphere with radius $R$, and
\begin{eqnarray}
    V = \frac{V^1_{D-2} R^{D-1}}{D-1},\nonumber
\end{eqnarray}
is the corresponding volume, with 
\begin{eqnarray}
 V^1_{D-2} =\frac{2 \pi^{(D-1)/2} }{ \Gamma[(D-1)/2]},\nonumber
 \end{eqnarray}
 being the volume of a unit $(D-2)$-sphere. The work density is defined as $W = - h^{ab} T_{ab} / 2$, where $h^{ab}$ and $T_{ab}$ are the projection of the metric and the energy momentum tensor into the two-dimensional normal space $(M^2, h_{ab})$, respectively, and the energy supply vector is given by $\Psi_\alpha = T_\alpha^b \partial_b \mathbf{R} + W \partial_\alpha \mathbf{R}$. The right-hand side of the unified first law can be rewritten in coordinates $(t, r)$ as
\begin{eqnarray}\label{308}
    \mathbf{A}\Psi_a dx^a + WdV = A(t,r)dt + B(t,r)dr.
\end{eqnarray}
Using (\ref{205}), we can get the $A(t,r)$ and $B(t,r)$ as follows
\begin{eqnarray}\label{309}
    A(t,r) &=&\mathbf{A}(T_t^{\ r} \mathbf{R}_{,r} - T_r^{\ r} \mathbf{R}_{,t}),\nonumber\\
    B(t,r) &=&\mathbf{A} (T_t^{\ r} \mathbf{R}_{,t} - T_t^{\ t}\mathbf{R}_{,r}),
\end{eqnarray}
with $\mathbf{R} = a r$. For the generalized Misner-Sharp energy to be well-defined, the expression $A(t,r) dt + B(t,r) dr$ must be integrable, which requires the following integrability condition to hold
\begin{eqnarray}\label{310}
    \frac{\partial A(t,r)}{\partial r} - \frac{\partial B(t,r)}{\partial t} = 0.
\end{eqnarray}
In the context of modified gravity theory, this condition is not always satisfied, for instance see the case of $f(R)$ gravity discussed in \cite{Cai:2006rs}. Here, assuming that integrability condition holds, we can integrate the unified first law to obtain the effective Misner-Sharp energy \(E_{eff}\). Hence, in $D$-dimensional FLRW spacetime, Misner-Sharp energy can be found as
\begin{eqnarray}\label{311}
    E_{eff} &=& \frac{\pi^{\frac{1}{2}(D-1)} (r a)^{D-1} }{2 D \kappa a^4 \Gamma\left[\frac{1+D}{2}\right]} \left[ -2 D \Lambda a^4\right. (D-2)(D-1) D a^2 \left( k + \dot{a}^2 \right) + (D-1) \kappa\sigma  \left( k + \dot{a}^2 \right)^2\nonumber\\
&&- \kappa(1-D) \left[ 4(D-1)\alpha + D\beta \right]  \big\{ k^2 (D-2)^2 - (D^2-4) \dot{a}^4 \nonumber\\
&&- D a^2 \ddot{a}^2 + 2 \dot{a}^2 \left( -2k (D-2) + (D-3)D a \ddot{a} \right)+  \left.2 D a^2 \dot{a} \dddot{a} \big\} \right].
\end{eqnarray}
This expression reduces to the standard Misner-Sharp energy in Einstein gravity when $\alpha ,\beta,\gamma= 0$ and \( D=4\) \cite{Cai:2009qf,Akbarieh:2025vau}. For the spatially flat universe, $k=0$, it takes the form
\begin{eqnarray}\label{312}
     E_{eff} &=& \frac{\pi^{\frac{1}{2}(D-1)} (r a)^{D-1} }{2 D \kappa a^4 \Gamma\left[\frac{1+D}{2}\right]} \big[ -2 D \Lambda a^4+\dot{a}^4(D-1)\sigma \kappa + a^2 \dot{a}^2 (D-2)(D-1)D\nonumber\\
  &&-(1-D)\left[ 4(D-1)\alpha + D\beta \right]\{ -a^2 \ddot{a}^2 D+ 2a^2 \dot{a} \dddot{a}D \nonumber\\
  &&+2 a \dot{a}^2 \ddot{a}(D-3)D - \dot{a}^4 (D^2-4)\}\kappa 
  \big].
\end{eqnarray}
In the asymptotic limit where the scale factor $a$ becomes exponentially large, corresponding to a de Sitter-like expansion dominated by the cosmological constant $\Lambda$, the generalized Misner-Sharp energy simplifies to 
\begin{eqnarray}\label{312}
     E^{(s)}_{eff}&=&\frac{V\Lambda}{4\kappa }(D-1)\big(D(D-3)+\kappa\Lambda(D-1)(D-4)[(D-1)(D\alpha+\beta)\nonumber\\
    &&+(D-3)(D-2)\gamma]\big),
\end{eqnarray}
which $E^{(s)}_{eff}$ is called the “saturated Misner-Sharp energy".
It is worth mentioning that the quadratic curvature terms, proportional to $\alpha$, $\beta$, and $\gamma$, in this limit, contribute to the Misner-Sharp energy. 
On the other hand, in the early universe, where the scale factor $a$ is small and the expansion rate $\dot{a}$ is large, the quadratic curvature terms significantly increase the Misner-Sharp energy due to the high curvature. In this regime, terms such as $\dot{a}^4/a^4$ or $\ddot{a}^2/a^2$ can dominate, leading to substantial deviations from the GR predictions. In addition, it can be concluded from (\ref{312}) that the parameter $\sigma$ makes an important contribution to the Misner-Sharp energy in higher dimensions. From (\ref{206}), for $D=4$, $\sigma$ goes to zero, reducing the contribution of quadratic curvature and aligning $E_{\text{eff}}$ with general relativity. In contrast, for $D>4$, a non-zero $\sigma$ enhances the energy in high-curvature regimes, potentially changing the cosmological dynamics. These behaviors suggest that quadratic gravity significantly affects the cosmological dynamics in higher dimensions, both in the early universe and late time depending on the curvature and coupling parameters \cite{Gialamas:2022xtt,deMedeiros:2025fqz}.

\section{Critical Points}\label{sec4}
\begin{figure}
    \centering
    \includegraphics[width=0.9\linewidth]{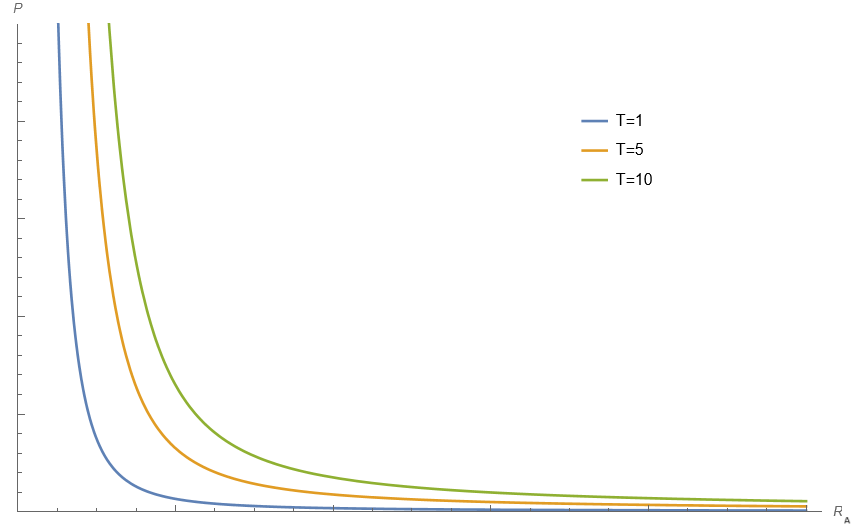}
    \caption{Pressure \(P\) as a function of \(R_A\) for \(\alpha = 1\) and \(\gamma = 1\) }
    \label{fig:401}
\end{figure}

\begin{figure}
    \centering
    \includegraphics[width=0.9\linewidth]{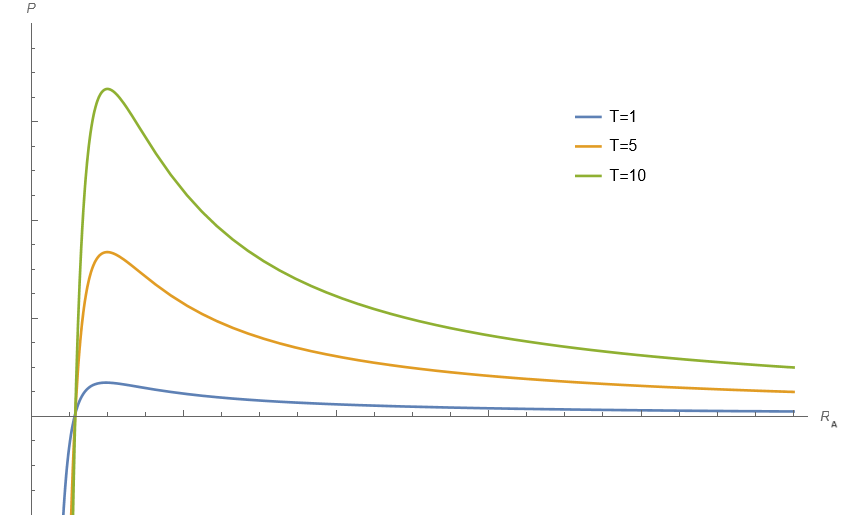}
    \caption{Pressure \(P\) as a function of \(R_A\) for \(\alpha = -1\) and \(\gamma = 1\).}
    \label{fig:402}
\end{figure}

\begin{figure}
    \centering
    \includegraphics[width=0.9\linewidth]{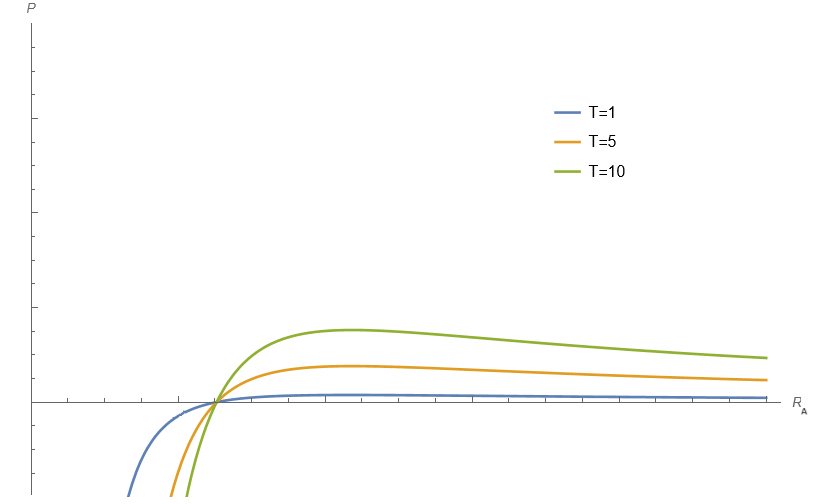}
    \caption{Pressure \(P\) as a function of \(R_A\) for \(\alpha = -1\) and \(\gamma = -1\) }
    \label{fig:403}
\end{figure}

In this section, we study the critical points of the thermodynamic system describing a $ D$-dimensional FLRW universe in the framework of quadratic gravity. Critical points are essential for understanding phase transitions, as they represent the conditions under which the system undergoes significant changes in its thermodynamic behavior. We use the thermodynamic equation of state to identify these critical points and investigate how higher-order curvature terms affect the phase structure \cite{Abdusattar:2022bpg}. To achieve this, we focus on the apparent horizon, which serves as a key geometric feature reflecting cosmological thermodynamics. The radius of the apparent horizon, $R_A$, captures the interaction between the Hubble expansion and the spatial curvature \cite{Tian:2014sca,Abdolmaleki:2015iaa}, which is modulated by the second-order gravitational corrections introduced in the Misner-Sharp energy. By calculating $R_A$, we can extract the Hawking temperature and the thermodynamic pressure, which enable us to construct the equation of state and identify critical points associated with phase transitions in the system. The apparent horizon in a $D$-dimensional FLRW universe is given by \cite{Akbar:2006kj}
\begin{eqnarray}\label{413}
    R_A = \frac{1}{\sqrt{H^2+k/a^2}},
\end{eqnarray}
where $H = \dot{a}/a$ is the Hubble parameter. The surface gravity at the apparent horizon is
\begin{eqnarray}\label{414}
\kappa_g= -\frac{1}{R_A}\left( 1-\frac{\dot{R_A}}{2 H R_A}\right),
\end{eqnarray}
and one can obtain the Hawking temperature as follows \cite{Cognola:2013fva}
\begin{eqnarray}\label{415}
   T=\frac{\left| \kappa_g \right|}{2\pi}.
\end{eqnarray}
By comparing the unified first law \cite{Cai:2006rs}
\begin{eqnarray}\label{419}
    dE=-TdS+WdV,
\end{eqnarray}
where $S$ represents the entropy, with the standard thermodynamic first law
\begin{eqnarray}\label{420}
    dU=TdS-PdV,
\end{eqnarray}
one identifies
\begin{eqnarray}\label{421}
U=-E, \nonumber\\
P= W.\;\;\;     
\end{eqnarray}
Using (\ref{204}) and (\ref{205}), the work density defined as
\begin{eqnarray}\label{416}
  W = -\frac{1}{2} h_{ij}T^{ij} =\frac{1}{2}(\rho - p),
\end{eqnarray} 
quantifies the thermodynamic work associated with the perfect fluid, where $\rho$ is the energy density $p$ and is the pressure. As we explained in the previous section, we can eliminate the contribution of the higher derivative of the scale factor by the following choice \cite{Gurses:2024tka}
\begin{eqnarray}\label{417}
    \beta = \frac{4(D-1)}{D} \alpha,
\end{eqnarray}
which constrains the quadratic gravity parameters. Using the metric (\ref{204}) and the apparent horizon (\ref{413}), the energy density and pressure of the perfect fluid are
\begin{eqnarray}\label{418}
    \rho &=&-\frac{\Lambda}{\kappa}+\frac{D-1}{2\kappa DR_A^4}\left( \kappa\sigma+D(D-2)R
    _A^2\right),
    \nonumber\\
    p&=&\frac{\Lambda}{\kappa} - \frac{1}{2 \kappa D R_A^4}\big[ D(D-2)R_A^2(D+8\pi R_A T -5 ) \nonumber\\
    && + (D+ 16 \pi R_A T -9)\kappa \sigma\big].
\end{eqnarray}
Thus, the thermodynamic pressure of (\ref{421}), (\ref{416}) and (\ref{418}) is
\begin{eqnarray}\label{422}
P(R_A,T)=-\frac{\Lambda}{\kappa} + \frac{1}{2\kappa DR_A^4}\big[(D+8\pi R_A T-5)  \kappa \sigma+D(D-2)(D+4\pi R_AT -3)R_A^2 \big].\nonumber\\
\end{eqnarray}
To find critical points, we apply the conditions \cite{Bena:2012hf,Hu:2018qsy,Arcadi:2021mag,Abdusattar:2023hlj}
\begin{eqnarray}\label{423}
    \Big(\frac{\partial P}{\partial R_A}\Big)_T = 0,
\end{eqnarray}
\begin{eqnarray}\label{423.5}
    \Big(\frac{\partial^2 P}{\partial R_A^2}\Big)_T = 0.
\end{eqnarray}
Solving the first condition (\ref{423}) gives the critical temperature
\begin{eqnarray}\label{424}
    T_c=-\frac{D(D-3)(D-2)R_c^2+2(D-5)\kappa\sigma}{2\pi R_c \left(6\kappa\sigma+D(D-2)R_c^2\right)},
\end{eqnarray}
Using (\ref{423.5}) and (\ref{424}), the critical radius $R_c$ can be obtained as
\begin{eqnarray}\label{425}
    R_{c}^{(\pm)}= \sqrt{\kappa\frac{-14(D-2) \pm  2 \sqrt{241+4D(13D-55)}}{(D-3)(D-2)D}\sigma}.\nonumber\\
\end{eqnarray}
Hence, there are two possible values for the critical radius, $R^{(\pm)}_c$. These values are only physically acceptable if they are real and positive. In dimensions $D=4$, and if $\sigma=0$ or correspondingly
 \begin{eqnarray}\label{426}
    \alpha = -\frac{D(D-3)}{(D-2)(D-1)} \gamma,
\end{eqnarray} 
phase transitions only occur at the beginning of the universe, i.e. at \(R^{(\pm)}_c=0\), and after that the universe remains in a steady state. Of course, this result seems obvious according to the \cite{Gurses:2024tka}, because if condition the \eqref{426} holds, our model will reduce to GR. If $\sigma<0$ or equivalently
\begin{eqnarray}
    \alpha < -\frac{D(D-3)}{(D-2)(D-1)} \gamma,
\end{eqnarray}
we find that $R_c^+$ equals zero at $D=5$, and is non-physical for other values of $D$, wheres $R_c^-$ is physical for $D>4$. If $\sigma>0$ or equivalently
\begin{eqnarray}
    \alpha > -\frac{D(D-3)}{(D-2)(D-1)} \gamma,
\end{eqnarray}
$R_c^+$ is physical for $D>4$, but $R^-_c$ does not exist for any values of $D$.\\
To better understand the thermodynamic behavior of this system, we investigate the pressure \(P\) as a function of the apparent radius \(R_A\) for different temperatures and model parameters. The diagrams (\ref{fig:401}), (\ref{fig:402}) and (\ref{fig:403}) show this dependence.
\\
Figures (\ref{fig:401}), (\ref{fig:402}) and (\ref{fig:403})  depict the thermodynamic pressure ($P$) as a function of the apparent horizon radius ($R_A$) for varying quadratic gravity parameters $\alpha$ and $\gamma$ in a $D$-dimensional FLRW universe. In Fig. (\ref{fig:401}) ($\alpha = 1$, $\gamma = 1$), the pressure exhibits exponential decay with increasing $R_A$ while remaining positive across all temperatures, suggesting thermodynamic instability due to the absence of a stable equilibrium state. In Fig. (\ref{fig:402}) ($\alpha = -1$, $\gamma = 1$), the pressure is initially negative at small $R_A$, indicating thermodynamic stability. As $R_A$ increases, the pressure rises, crosses zero, and becomes positive, signaling a transition to thermodynamic instability at larger scales. Finally, the pressure asymptotically approaches zero from above. In Fig. (\ref{fig:403}) ($\alpha = -1$, $\gamma = -1$), the pressure is initially negative at small $R_A$; As $R_A$ increases, the pressure crosses zero and becomes positive for higher temperatures, reflecting a transition to thermodynamic instability at larger scales. Therefore, the sign of the parameters $\alpha$ and $\gamma$ significantly affects the thermodynamic behavior: negative values tend to ensure stability at small scales, while the behavior at large scales depends on temperature and can lead to instability.\\
Note that the expression $D(D-2)(D-3)$ appears in the denominator of the critical radius relation (\ref{425}). At $D = 3$, this expression vanishes, rendering the critical radius formally divergent, which physically indicates the absence of any phase transition in three dimensions; this is consistent with physical expectations from gravitational models in low dimensions, since at $D=3$ gravity lacks dynamical degrees of freedom. Furthermore, in the limit $\sigma \to 0$, which is equivalent to approaching a specific condition of dynamical equilibrium, the critical radius tends to zero. This means that any phase transitions occur only on very small scales, i.e. at the beginning of the universe when the apparent horizon radius is small and the curvature of spacetime is large. In this region, especially when the condition $\sigma \to 0$ holds, which corresponds to certain values of the coupling parameters, the critical radius tends to zero. Consequently, after this initial stage and as $R_A$ becomes large, the system remains in a thermodynamically stable state and its behavior becomes similar to general relativity, in which no further phase transitions occur. In physical terms, this result shows that the effect of the quadratic curvature corrections is only significant in the early universe, and that these effects gradually disappear as the universe expands and the curvature decreases. In contrast, if $\sigma$ is large (e.g., due to strong gravity or strong non-minimal coupling), the critical radius can become very large, indicating that the system remains thermodynamically unstable over an extended period of cosmic time and only experiences phase transitions after a long evolutionary timescale.\\
The obtained \(R^{(\pm)}_c=0\) clearly shows a bifurcation structure that leads to the existence of two different branches for the critical radius, each of which may have a physical or unphysical behavior depending on the values of $\alpha$, $\gamma$, and $D$. An important aspect of this structure is that if the parameter $\alpha$ passes the critical value (\ref{426}), a branch of the critical radius $R^{(-)}_c$ jumps from zero to a positive value; this represents a structural transition from a stable to an unstable state. Such behavior, from the perspective of chaos theory and system dynamics, indicates the occurrence of a bifurcation point in the parameter space. In this framework, the model can be considered an effective dynamical system with several critical points that depend on the coupling parameters. Changes in these parameters lead to the appearance or disappearance of these equilibrium points and consequently to a qualitative change in the thermodynamic behavior of the system.

\section{Thermodynamic Properties}\label{sec5}
\begin{figure}
    \centering
    \includegraphics[width=0.9\linewidth]{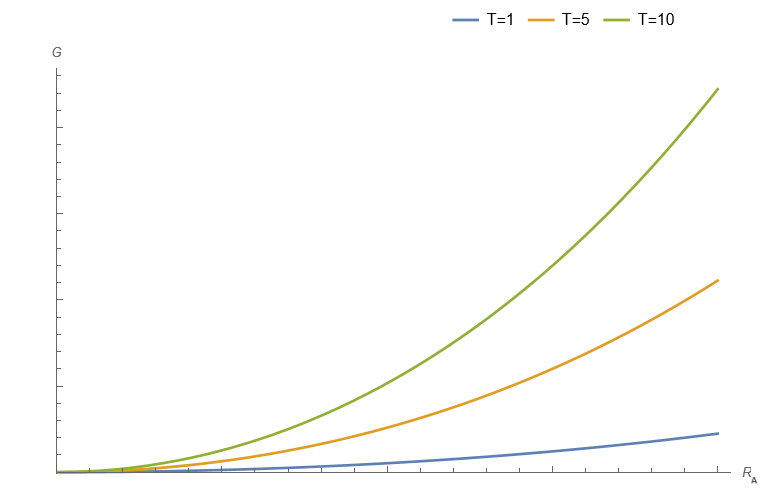}
    \caption{Gibbs free energy \(\mathbf{G}\) as a function of \(R_A\)  for \(\alpha = 1\) and \(\gamma = 1\)}
    \label{fig:501}
\end{figure}

\begin{figure}
    \centering
    \includegraphics[width=0.9\linewidth]{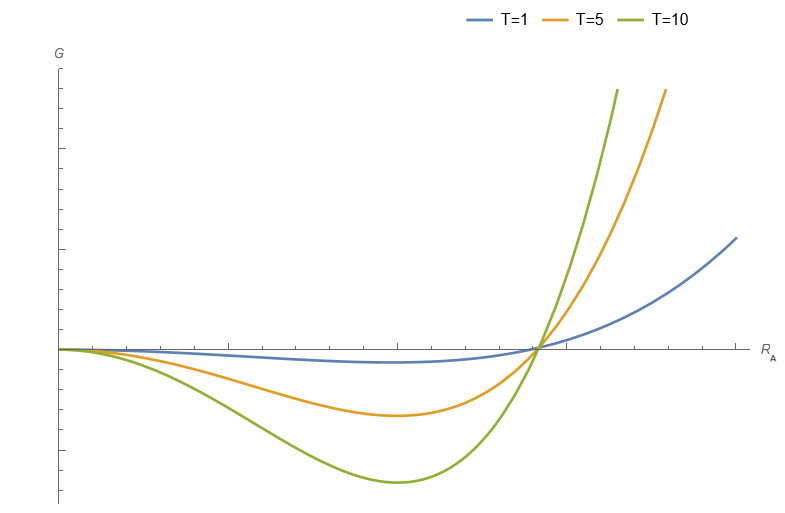}
    \caption{Gibbs free energy \(\mathbf{G}\) as a function of \(R_A\) for \(\alpha = -1\) and \(\gamma = 1\)}
    \label{fig:502}
\end{figure}

\begin{figure}
    \centering
    \includegraphics[width=0.9\linewidth]{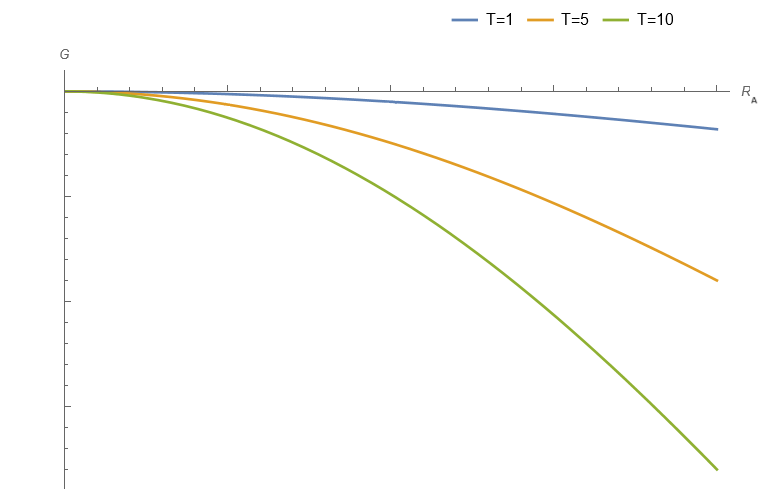}
    \caption{Gibbs free energy \(\mathbf{G}\) as a function of \(R_A\) for \(\alpha = -1\) and \(\gamma = -1\)}
    \label{fig:503}
\end{figure}

In this section, we investigate the thermodynamic properties of the quadratic gravity model in the $D$-dimensional FLRW background. Using the generalized Misner-Sharp energy and the equation of state obtained in the previous section, we calculate thermodynamic quantities, including Wald entropy, specific heat at constant pressure, enthalpy, Gibbs free energy. The equation of state is given as
\begin{eqnarray}
    P(R_A,T)=C(R_A)T+D(R_A),
\end{eqnarray}
where $C(R_A)$ and $D(R_A)$ are functions of the apparent horizon radius $R_A$. This form of the equation of state is inspired by the first law of thermodynamics at the horizon, in order to separate the temperature-dependent and temperature-independent contributions to the pressure. This separation allows for easy calculation of other thermodynamic quantities such as entropy via Legendre transformations \cite{Hansen:2016gud,Yang:2014kna}. The Wald entropy can be obtained as follows
\begin{eqnarray}\label{526}
    S = \int{V'(R_A) C(R_A) dR_A},
\end{eqnarray}
where $V'(R_A)$ represents the derivative of the volume with respect to $R_A$, and $C(R_A)$ is obtained from the equation of state as \cite{Zheng:2019mvn}
\begin{eqnarray}\label{427}
    C(R_A)=\frac{1}{D\kappa R_A^3}(4\pi\kappa \sigma+2D(D-2)\pi R_A^2).
\end{eqnarray}
By integrating (\ref{526}), we have
\begin{eqnarray}\label{428}
S=\frac{4 \pi^{(D+1)/2}}{D\Gamma[(D-1)/2]}R_A^{D-4}(\frac{DR_A^2}{\kappa}+\frac{2\sigma}{D-4}).
\end{eqnarray}
The above relation shows that the entropy has a direct dependence on the parameter $\sigma$, which is due to the presence of quadratic terms in the gravitational action. In particular, the contribution of $\sigma$ in the expression $2\sigma/(D-4)$ is not independent of the apparent horizon radius $R_A$, but is amplified by the power of $R_A^{D-4}$, especially in higher dimensions. This clearly shows that the effect of the second-order curvature terms on entropy is much more significant in higher-dimensional spacetimes.
The enthalpy, $\mathcal{H} = -E_{eff} + P V$, is derived as
\begin{eqnarray}\label{430}
    \mathcal{H} =\frac{\pi^{(D-1)/2}}{D \kappa\Gamma[(D+1)/2]}  R_A^{D-5} (2 \pi R_A T - 1 ) ( D(D-2)R_A^2 + 2 \kappa\sigma ).
\end{eqnarray}
This expression captures the interplay between the FLRW universe's geometry (via $R_A$ and $D$) and thermodynamic variables (via $T$ and $\sigma$), reflecting the total energy content, including contributions from quadratic gravity terms.
Next, we compute the specific heat at constant pressure, $C_P$, which indicates the system's thermodynamic stability. It is given by
\begin{eqnarray}\label{429}
    C_P&=& \left( \frac{\partial \mathcal{H}}{\partial T} \right)_P=
    \frac{2 \pi^{(D+1)/2}R_A^{D-4}(D(D-2)R_A^2+2\kappa\sigma)}{\Gamma[(D+1)/2]D\kappa} \nonumber \\
     && \times \frac{D(D-2)(3-D+2(D-2)\pi R_AT)R_A^2 + 2(5-D+2(D-4)\pi R_A T)\kappa\sigma}{D(D-2)(D-3+2\pi R_A T)R_A^2 +2(D-5+6\pi R_A T)\kappa \sigma }.\;\;\;\;\;\;\;\;\;\;\;
\end{eqnarray}
The analysis of the $ C_P $ reveals that its sign serves as a critical indicator of the system's thermodynamic stability. Specifically, positive values of $ C_P $ are indicative of thermodynamic stability, whereas negative values may suggest the occurrence of phase instability. The influence of the coupling parameters $ \alpha $ and $ \gamma $, which are incorporated into the equation through $ \sigma $, plays a more prominent role in higher-dimensional spacetimes ($ D > 4 $) and can lead to qualitative shifts in the system's phase structure. In the specific case of $D=4$, the vanishing of $\sigma$ eliminates the contribution of second-order curvature terms, resulting in convergence toward the predictions of general relativity. Numerical evaluation of this equation with varying values of $ \alpha $ and $ \gamma $ demonstrates that thermodynamic instability may emerge in high-curvature regimes (characterized by small $ R_A $ values), while an increase in $ R_A $ can promote stability. These findings align with the observed behaviors of the Gibbs free energy (as depicted in Figures (\ref{fig:501}) to (\ref{fig:503})) and underscore the significant role of quadratic gravity corrections in shaping the early cosmological dynamics. One of the important points that has received less attention in the thermodynamic analysis of the quadratic gravity model in the context of FLRW is the dependence of the thermodynamic behavior on the dynamical structure of the apparent horizon over cosmological time \cite{Rudra:2020rhs}. Given that $R_A$ itself is a function of cosmic time, it can be concluded that the transition between regions with positive and negative Gibbs energy may lead to temporal transitions between stable and unstable phases in the dynamical history of the universe. In other words, the universe may be in a thermodynamically unstable phase in the early stages (when $R_A$ is small) and, with the passage of time and the expansion of the horizon, enter a region of parameters that exhibit stable behavior. This interpretation can provide a framework for understanding phenomena such as the spontaneous stabilization of the universe in generalized gravitational theories \cite{Csaki:2023pwy}.
\\
The Gibbs free energy, $\mathbf{G} = \mathcal{H} - T S$, is
\begin{eqnarray}\label{431}
\mathbf{G} &=&-\frac{\pi^{(D-1)/2}}{D(D-4)\kappa\Gamma[(D+1)/2] }  R_A^{D-5}(D(D-4)(D+2\pi R_AT-2)R_A^2\nonumber \\
  && + 2(D+6\pi R_A T -4)\kappa \sigma ).
\end{eqnarray}
Based on the analysis of the graphs in Figures (\ref{fig:501}), (\ref{fig:502}), and (\ref{fig:503}), the behavior of Gibbs free energy ($\mathbf{G}$) in the $D$-dimensional FLRW universe and the framework of quadratic gravity has been investigated. The results clearly show that the free energy exhibits different behaviors under the influence of the quadratic gravity parameters $\alpha$ and $\gamma$. In the case of $\alpha=1$ and $\gamma=1$ (Figure \ref{fig:501}), the Gibbs energy $\mathbf{G}$ increases with the increase in the apparent horizon radius $R_A$ and has a positive value at all temperatures examined ($T = 1, 5, 10$). This behavior indicates the thermodynamic instability of the system, because a positive Gibbs free energy means that the system is not in a thermodynamically stable state. In the case of $\alpha=-1$ and $\gamma=1$ (Figure \ref{fig:502}), initially, as $R_A$ increases, the value of $\mathbf{G}$ decreases and remains negative, indicating thermodynamic stability at small scales. However, after passing a certain value of $R_A$, $\mathbf{G}$ becomes positive, indicating loss of stability at larger scales. This case shows that the thermodynamic behavior of the system is horizon scale dependent. In the case of $\alpha=-1$ and $\gamma=-1$ (Fig. \ref{fig:503}), the Gibbs energy $\mathbf{G}$ always decreases with increasing $R_A$ and remains negative in all intervals. This indicates the complete thermodynamic stability of the system at all scales and temperatures considered. In general, it can be concluded that the quadratic gravity parameters ($\alpha$ and $\gamma$), especially their sign, play an important role in determining the thermodynamic behavior of the system. Negative values of $\mathbf{G}$ mean stability conditions,while positive values indicate instability. Therefore, unlike Einstein gravity, which does not exhibit such features, the presence of terms with second-order derivatives of curvature has a significant impact on the thermodynamic behavior of the universe and can even lead to new unstable or stable phases in the multidimensional universe.

Also, since the $\sigma$ contribution appears in all thermodynamic quantities, this parameter can play the role of a thermodynamic pseudo-charge, whose effect is similar to that of electric charges in black hole thermodynamics \cite{MahdavianYekta:2020lde}. In this view, $\sigma$ can be considered as a new degree of freedom that leads to a different phase structure than Einstein gravity, especially in higher-dimensional spacetimes that are more sensitive to quadratic terms. A more detailed investigation of this phase structure, including the study of critical points and the thermal behavior near them, can provide new perspectives for analyzing the thermal dynamics of the universe.

\section{Conclusion}\label{sec6}
In this paper, we investigate the thermodynamic properties of the quadratic gravity model in the $D$-dimensional FLRW spacetime. First, by adding second-order curvature terms to the action, we derive the modified field equations and the effective energy-momentum tensor. Then, we calculate the generalized Misner-Sharp energy in this model and show that in the absence of quadratic terms, the results revert to GR.\\
Next, we examine the thermodynamic behavior of the apparent horizon \eqref{413} and derive the thermodynamic equation of state \eqref{422}, that is, the relation among pressure, temperature and the radius of the apparent horizon. Using this equation, we obtain the critical points \eqref{425} and critical temperature of the system \eqref{424}, demonstrating that quadratic terms have a significant impact on the phase structure and thermodynamic stability of the model. Moreover, we investigate the role of the parameter $\sigma$, which depends on the couplings $\alpha$, $\beta$, and $\gamma$, in determining whether stable or unstable phases occur. Several thermodynamic quantities, such as Wald entropy, specific heat, enthalpy, and Gibbs free energy, are also calculated. The results indicate that under specific conditions, as time passes, the system experiences a transition from an initially unstable phase to a stable one.\\
Furthermore, we emphasize that the apparent horizon radius, as a time-dependent dynamical quantity, serves as a useful indicator in tracking the transition between thermodynamic phases. This reflects the fact that the universe may evolve from an unstable early phase into a thermodynamically stable regime as a consequence of its underlying dynamics, with the horizon radius capturing this evolution.



\end{document}